\documentclass[useAMS,usenatbib,aps,floatdix,nofootinbib]{mn2e}
\usepackage{graphics}
\usepackage{graphicx}
\usepackage{amssymb,amsmath}
\usepackage{amsmath}
\usepackage{psfrag}
\usepackage{graphicx}

\usepackage{color}

\newcommand{\gsc}{\gamma_{\rm sc}}

\newcommand{\f}{\frac}

\newcommand{\be}{\begin{equation}}      
\newcommand{\ee}{\end{equation}}      
      
\newcommand{\bef}{\begin{figure*}}      
\newcommand{\eef}{\end{figure*}}      
\newcommand{\bea}{\begin{eqnarray}}    
\newcommand{\eea}{\end{eqnarray}}      
  
\def\spose#1{\hbox to 0pt{#1\hss}}
\def\ltapprox{\mathrel{\spose{\lower 3pt\hbox{$\mathchar"218$}}
\raise 2.0pt\hbox{$\mathchar"13C$}}}
\def\gtapprox{\mathrel{\spose{\lower 3pt\hbox{$\mathchar"218$}}
\raise 2.0pt\hbox{$\mathchar"13E$}}}
\def\inapprox{\mathrel{\spose{\lower 3pt\hbox{$\mathchar"218$}}
\raise 2.0pt\hbox{$\mathchar"232$}}}

\def\bse{\begin{subequations}}
\def\ese{\end{subequations}}

\def\bF{{\mathbf F}}

\def\lsim{\raise 0.4ex\hbox{$<$}\kern -0.8em\lower 0.62ex\hbox{$\sim$}} 
\def\gsim{\raise 0.4ex\hbox{$>$}\kern -0.7em\lower 0.62ex\hbox{$\sim$}}

\def\f0N{f_0^{(N)}}
\def\bec{\begin{center}}
\def\eec{\end{center}}

\title[Resolution of cosmological N-body simulations] 
{Stable clustering and the resolution of dissipationless
  cosmological N-body simulations}
\author[D. Benhaiem, M. Joyce and F. Sylos Labini]{David
Benhaiem${^{1,2}}$, Michael Joyce${^{2,3}}$ and Francesco Sylos
Labini${^{4,1,5}}$\\ $^{1}$ Istituto dei Sistemi Complessi Consiglio
Nazionale delle Ricerche, Via dei Taurini 19, 00185 Rome,
Italy\\ $^{2}$UPMC Univ Paris 06, UMR 7585, LPNHE, F-75005, Paris,
France\\ $^{3}$CNRS IN2P3, UMR 7585, LPNHE, F-75005, Paris, France
\\ $^{4}$Museo Storico della Fisica e Centro Studi e Ricerche ``Enrico Fermi'', Via Panisperna 89 A,
00184 Rome, Italy 
\\ $^5$INFN Unit Rome 1,
Dipartimento di Fisica, Universit\'a di Roma Sapienza, Piazzale Aldo
Moro 2, 00185 Roma, Italy }

\begin{document}

\date{\today}

\maketitle

\begin{abstract}
The determination of the resolution of cosmological N-body
simulations, i.e., the range of scales in which quantities measured in
them represent accurately the continuum limit, is
an important open question.  We address it here using scale-free
models, for which self-similarity provides a powerful tool to control
resolution.  Such models also provide a robust testing ground for the
so-called stable clustering approximation, which gives simple
predictions for them.  Studying large N-body simulations of such
models with different force smoothing, we find that these two issues
are in fact very closely related: our conclusion is that the accuracy
of two point statistics in the non-linear regime starts to degrade 
strongly around the scale at which their behaviour deviates
from that predicted by the stable clustering hypothesis. Physically 
the association of the two scales is in fact simple to understand: stable 
clustering fails to be a good approximation when there are strong 
interactions of structures (in particular merging) and it is precisely 
such non-linear processes which are sensitive to fluctuations at the 
smaller scales affected by discretisation. 
Resolution may be further 
degraded if the short distance gravitational smoothing scale is larger
than the scale to which stable clustering can propagate.
We examine in detail the very different conclusions of studies by 
\cite{smith2003stable} and \cite{widrow_etal2009} and  find that the  
strong deviations from stable clustering reported by these works are the results of over-optimistic 
assumptions about scales resolved accurately by the measured 
power spectra, and the reliance on Fourier space analysis. We emphasise 
the much poorer resolution obtained with the  power spectrum compared 
to the two point correlation function.
 \end{abstract}


%%%%%%%%%%%%%%%%%%%%%%%%%%%%%%%%%%%%%%%%%%%%%%%%%%%%%%%%%%%%%%%%%%%%
%%%%%%%%%%%%%%%%%%%%%%%%%%%%%%%%%%%%%%%%%%%%%%%%%%%%%%%%%%%%%%%%%%%%

\section{Introduction} % (fold)
\label{sec:introduction}

Numerical simulations using the N-body method are the primary
instrument used to probe the non-linear regime of structure formation
in cosmology and provide the basis for all theoretical predictions for
the distribution of dark matter at the corresponding physical scales.
Over the last few decades, such simulations have gained in refinement
and complexity and have allowed the exploration of an ever larger
range of scales (for a review see e.g. \cite{bertschinger_98,springel2005simulations,dehnen+reed_2011}). Nevertheless, the
understanding of their precision and their convergence toward the
continuum limit remains, at very least, incomplete, in particular for
smaller scales (see e.g., \cite{splinter_1998, knebe_etal_2000,
  romeo_etal_2008, discreteness3_mjbm,power2016spurious}).  In this
context ``scale-free" cosmological models, in which both the expansion
law and the power spectrum characterizing the initial fluctuations are
simple power laws, have the advantage of relative simplicity, and they
have for this reason been studied quite extensively in the literature
(see e.g.
~\cite{efstathiou1988gravitational,colombi_etal_1996,bertschinger_98,jain+bertschinger_1998,smith2003stable,knollmann_etal_2008,
widrow_etal2009,orban2013keeping,diemer+kravtsov_2015}).  More
specifically these models provide a testing ground for the numerical
method through the predicted ``self-similarity" of the clustering: the
temporal evolution of the clustering statistics must be equivalent to
a rescaling of the distances. This follows from the fact that there is
only one characteristic length scale (derived from the amplitude of
the fluctuations) and one characteristic time scale in the
model. Further the exact rescaling function can be determined from the
evolution in the linear regime of arbitrarily small
fluctuations. However, discreteness and numerical effects typically
introduce additional characteristic scales (e.g., force regularization
at small scales, particle density, finite box size, etc.) which
lead directly to a breaking of such self-similarity. Thus the self
similarity of clustering provides a potentially powerful tool to
separate the scales affected by such non-physical effects from the
physical results representing the continuum limit. The focus of this
study is to exploit self-similar models to better understand the
resolution at small scales of N-body simulations. In particular we
will use simulations with a very small force smoothing which allow us
to follow carefully the propagation of self-similarity to small scales
in the course of a simulation.

A further motivation for studying scale-free models is that they
provide a very simple analytical prediction for non-linear clustering
which is the stable clustering hypothesis
(\cite{davis+peebles_1977,peebles}).  This corresponds to the
assumption that once a structure is strongly non-linear it no longer
evolves in physical coordinates, i.e., structures behave as though
they were isolated virialized structures.  While this hypothesis can
be made in any cosmological model, for scale-free initial cosmologies
it implies, when combined with self-similarity, that the strongly
non-linear regime of the two point correlation function (and
also of the power spectrum) should be a power law function of
the separation, i.e. $\xi(x) \propto x^{-\gsc}$ where the exponent
$\gsc$ is a simple function of $n$, the exponent characterizing the
power law behaviour of the initial fluctuations (with power spectrum
$P(k) \sim k^n$). The stable clustering hypothesis can, at
best, be a good approximation because it neglects in principle 
the evolution of structures due to their interaction in general 
(and their merging in particular). It is, nevertheless, a fundamental 
question about non-linear clustering to understand how good an 
approximation stable clustering in fact provides. Indeed, the assumption 
of the validity of this approximation at sufficiently small scales provided 
the basis of the assumed functional form of non-linear clustering at small scales in
phenomenological approaches, like that of \cite{hamilton_etal_1991, peacock} 
(hereafter PD), which were widely used to compare galaxy data to 
cosmological models until a few years ago.

Historically there have been numerous numerical studies of the
validity of the stable clustering hypothesis in scale-free models,
with, for a long time, inconclusive results. While, for example,
 \cite{padmanabhan1995pattern} and \cite{colombi_etal_1996} 
reported deviations from stable clustering,  ~\cite{jain1997does},
\cite{bertschinger_98} and \cite{valageas_etal_2000}  found 
results  apparently in agreement with this hypothesis
in the  strongly non-linear regime. A subsequent
larger study, by \cite{smith2003stable}, reported
clear deviations from the stable  clustering predictions at smaller 
scales. These results, confirmed also by the 
larger study of \cite{widrow_etal2009}, appeared thus to
unambiguously detect the inadequacy of the stable clustering
hypothesis, and more specifically of the PD fits to the non-linear
clustering based on it. The latter have then been superseded by fits
with ``halo models" which generically break stable clustering. Indeed
these models are explicitly based on the assumption of smooth
virialized structures which are built up through merging, which is
qualitatively different from stable clustering which
instead implies a hierarchy of virialized structures~\footnote{Nevertheless 
it is possible, as shown in 
\cite{ma+fry_2000a}, to write down very specific halo models 
which have the exponents predicted by stable clustering at
asymptotically small scales.}.

In this paper we closely re-examine the issue of the breakdown of
stable clustering in scale-free cosmological models, which is, as we
will see, inseparable from the issue of the resolution of N-body
simulations of these models. We have been prompted to carry out the
simulations and analysis reported here by results we obtained using
smaller simulations, reported in a previous paper
\citep{benhaiem2013self} in which we explored clustering in 
scale free models in a broader class than usually considered in cosmology. 
The conclusions of this study appeared to be discrepant with those 
of \cite{smith2003stable}, which, as discussed above, have been 
widely assumed in the literature to establish definitively clear {deviations} 
of non-linear clustering from that predicted by the stable clustering 
hypothesis.  Indeed our conclusion --- using simulations somewhat 
smaller than those of \cite{smith2003stable}, but with higher resolution --- 
was that the resolved (i.e. self-similar) non-linear clustering was in good
agreement with the stable clustering hypothesis. Moreover, while
we observed apparent deviations from the stable clustering predictions
like those reported by \cite{smith2003stable}, these were not in the
self-similar regions.  Further we have detected a clear
  dependence on force smoothing $\varepsilon$, by comparing
  simulations with different $\varepsilon$, precisely in the range of
  scales which has been considered by \cite{smith2003stable} in their
  fits. This would imply that the assumptions made by
  \cite{smith2003stable}, when obtaining their fits to the power
  spectrum, are strongly affected by force smoothing. 
    
The results of \cite{smith2003stable} for non self-similar
 cosmologies, and notably the standard $\Lambda$CDM
cosmology, given in terms of the parameters
of the ``halofit'' model, have been very extensively used in the
literature. These have been revisited by other authors. In
  particular, \cite{takahashi_etal_2012} found that the
  results of 
\cite{smith2003stable} for the power spectrum at small scales
(large wave-numbers) are indeed incorrect, due to the
  underestimation of the power generated by the effect of
smoothing. \cite{takahashi_etal_2012} have corrected the halofit
power spectrum, and this change has then been widely
adopted in the literature. While the correction of the halofit
power spectrum was taken into account for the $\Lambda$CDM
simulations, the consequences for the results  of 
\cite{smith2003stable} for scale-free simulations,
and in particular for the issue of the validity of 
stable clustering, have not been examined in the literature
other than in one other study \citep{widrow_etal2009}.

In the present work we thus choose our simulations to allow a detailed
comparison with the results of \cite{smith2003stable} for scale-free
models, and to assess, in particular, the role of the force
smoothing length in limiting their resolution. Specifically we present the 
study of six simulations, for the cases $n=-2$, $n=-1$ and $n=0$, and
for $N=256^3$ particles (as \cite{smith2003stable}) and an 
Einstein-de Sitter (EdS) cosmology. For each case, we have run 
simulations with exactly the same initial conditions and numerical parameters,
changing only the force smoothing.  On the one hand we have used the same
smoothing used by \cite{smith2003stable}, and, on the other hand, a smoothing
as in \cite{benhaiem2013self}, smaller by a factor of six.
The detailed analysis  of these simulations allows
us to draw clear conclusions concerning the results of
\cite{smith2003stable} (and also \cite{widrow_etal2009})
and in particular the issue of the validity of stable clustering.  
It also reveals that there is in fact an intimate
connection between the breakdown of this same approximation and the
resolution of N-body simulations. Our study also allows us
to address in detail the important issue of optimal choice of
smoothing in a cosmological simulation.

The paper is organized as follows. We first recall, in
Sect.\ref{sec:a_family_of_scale_free_cosmologies}, the equations 
of motion in an expanding universe, the self-similar evolution 
of scale-free models and the prediction obtained in the stable 
clustering hypothesis for the two point correlation function. 
In Sect.\ref{Numerical_simulations} we describe the numerical 
simulations, and in Sect.\ref{sec:results} we present our results.
Finally in Sect.\ref{sec:stable_clustering} we summarise our
main conclusions.

%%%%%%%%%%%%%%%%%%%%%%%%%%%%%%%%%%%%%%%%%%%%%%%%%%%%%%%%%%%%%%%%%%%%%%%%
%%%%%%%%%%%%%%%%%%%%%%%%%%%%%%%%%%%%%%%%%%%%%%%%%%%%%%%%%%%%%%%%%%%%%%%%

\section{Scale-free cosmologies} % (fold)
\label{sec:a_family_of_scale_free_cosmologies}

 {In a scale-free cosmology both} the power spectrum,
 characterising initial matter density fluctuations, and the
 cosmological expansion are simple {power law functions}. In
 practice, the latter is usually taken {to be a EdS cosmology},
 with the scale factor
\begin{equation}
a(t)= \left (\frac{t}{t_0} \right)^{2/3} \quad {\rm where} \quad
t_0=\frac{1}{\sqrt{6\pi G \rho_0}}
\label{eds-scalefactor}
\end{equation}
and where $\rho_0$ is the mean mass density in comoving
coordinates. {A power spectrum} $P(k) \propto k^n$ with $n > -3 $
introduces (through its amplitude) a single length scale
$L_0$. {This scale can be defined} to be that at which {density}
fluctuations have, initially, {a given } amplitude (e.g.,
unity). 

{The gravitational evolution of such a universe} can, in principle,
be written in rescaled dimensionless variables (taking, e.g., $t_0$
and $L_0$ equal to unity); as a consequence any evolved quantity 
is a function only of these variables. 
In particular, any {two-point statistic} can be written as a
function of the form $f(x/L_0, t/t_0)$.  As the time $t_0$ is
arbitrary, one can then infer that the temporal evolution is
``self-similar", {which in this context means} equivalent to a
spatial rescaling. We thus have, for the two-point {correlation
  function (CF)},
\begin{equation}
	\xi(x, a)=\xi_0 \Big(\frac{x}{R_s(a)} \Big) \;, 
	\label{self-similarity-xi}
\end{equation}
where $\xi_0=\xi(x, 1)$ and $a_0=1$ is an arbitrary reference scale
factor. Analogously, the dimensionless {power-spectrum (PS) is}
defined by{
\[
\Delta^2 (k) = \frac{4\pi k^3}{\left( 2 \pi \right)^3} P(k)
\]
and, if the evolution is self-similar,}
we
have
\begin{equation}
	\Delta^2(k,a)=\Delta^2_0(k R_s(a))
	\label{self-similarity-delta}
\end{equation}
where the function $R_s(a)$ is defined as the scale $L_0$, but
for a scale factor $a$.

Finally the dependence of $R_s(a)$ on $a$ can be inferred from linear
theory: assuming that it describes the evolution of 
fluctuations at large enough scales, 
with $\Delta^2(k,a) \propto a^2$,
we can infer that
 \begin{equation}
	R_s = a^{\frac{2 }{3+n}}\,.
	\label{eq-RS_SS}
\end{equation}

One important remark on this derivation: the PS of fluctuations in a
perturbed FRW cosmology cannot, strictly, be a pure power law because
its integral is proportional to the one point variance of the density
fluctuations which must be finite (see e.g., \cite{book}). Implicitly
we have thus assumed in the above {derivation} that there is an
ultraviolet cut-off in the PS, and, crucially, that the gravitational
clustering does not depend on it.  Thus, in practice, only insofar as
any such dependence on this scale is wiped out by the dynamics, can
one expect the clustering at a given scale to be self-similar. On the
basis of general theoretical arguments given originally by Zeldovich
(see e.g., \cite{peebles}), it is very plausible that such an
assumption should be valid for exponents $n \le 4$. In practice
self-similarity has been observed numerically in models up to $n=2$
\citep{benhaiem2013self} in {the three dimensional space}, and up
{to} $n=4$ in analogous one dimensional {models}
\citep{joyce+sicard_2011,benhaiem2013exponents}.  In the N-body problem as 
discussed below, the ultraviolet cut-off to the power spectrum 
is provided by the Nyquist frequency of the initial perturbed 
lattice configuration: testing for self-similarity is, as we emphasize, a 
way to test that the evolved system does not depend on this 
unphysical scale.
power law

\subsection{Stable clustering prediction for two point statistics}

{According to the} stable clustering hypothesis {the
  statistical properties of non-linear clustering} are, { in
  physical coordinates}, time independent. This {is} true, to a
very good approximation, {if} matter in highly overdense regions
behaves as if it were isolated from the rest of {the universe}.
If we assume that this hypothesis {is} valid in a scale-free model, in
which  clustering {develops in a self-similar way}, {then}
there is no characteristic length scale. {Indeed,} any
characteristic scale would be proportional, in comoving coordinates,
to $1/a$ because of stability, which is in contradiction with the
supposed self-similar scaling. The only possibility left is then that
{CFs} are simple power laws. For the resultant two-point {CF},
{with a behaviour of the type} $\xi(x) \sim x^{-\gsc}$, it is
straightforward to calculate that \citep{davis+peebles_1977, peebles,
  benhaiem2013self}
\begin{equation}
	\gamma_{\rm sc} \left( {n } \right) = 
		\frac{3(3+n)}{5+n}\,
	\label{sc-prediction-Peebles}
\end{equation}
with $0 < \gsc < 3$ for $n>-3$. Thus, the larger the value of $n$, the
steeper is the predicted {exponent of the two-point CF}.

%%%%%%%%%%%%%%%%%%%%%%%%%%%%%%%%%%%%%%%%%%%%%%%%%%%%%%%%%%%%%%%%%%%%%%%%

\subsection{Physical meaning of stable clustering exponent}
\label{sub:the_relative_size_of_structures}

It is instructive for what follows to recall a simple relation
involving the stable clustering exponent $\gsc$ which we have derived
in \cite{benhaiem2013self} (see also \cite{benhaiem2013exponents}).
Let us follow the evolution of two initially overdense spherical
regions, of initial comoving size $L_1^0$ and $L_2^0$ respectively,
with $L_1^0 \ll L_2^0$.  Assuming scale free initial conditions,
the ratio of their initial rms density fluctuation is thus $(L_2^0/
L_1^0)^{(n+3)/2}$. Assuming further that they collapse and then
virialize at a time which depends only on the linearly extrapolated
amplitude of such an initial overdensity, we can infer that the ratio
of the corresponding scale factors (at which virialization occurs)
 \begin{equation}
\left(\frac{a_2}{a_1}\right)= 
	\left(\frac{L_2^0}{L_1^0}\right)^{\frac{3+n}{2}} \;.
\end{equation}
If we now assume that the smaller structure, which virializes first
(at scale-factor $a_1$), remains stable subsequently, we can infer
that the relative size of the two structures at $a_2$ is
\begin{equation}
 \frac{L_1}{L_2} = \frac{L_1^0}{L_2^0} \times
 \left(\frac{L_1^0}{L_2^0}\right)^{\frac{3+n}{2}} =
 \frac{L_1^0}{L_2^0} \times
 \left(\frac{L_1^0}{L_2^0}\right)^{\frac{\gsc}{3 - \gsc}}.
\label{Eq-scaling-structure}
\end{equation}
The multiplicative factor on the right hand side of this equation is
strictly less than unity, and quantifies the change in the {\it
relative size of two structures between the linear and strongly
  non-linear regime}, assuming stable clustering applies after
virialization.  In other words, the multiplicative factor quantifies
how much a structure which virializes ``shrinks" due to stable
clustering relative to a structure which collapses and virializes
later. 

For a fixed range of initial comoving scales, the reduction in
relative size increases as $\gamma_{sc}$ does (or correspondingly as
$n$ increases). The greater is $\gamma_{sc}$ the more ``concentrated''
are the pre-existing virialized substructures inside a larger
structure when it collapses.  It is in fact precisely such a
``relative shrinking" which one might expect to make stable clustering
a reasonable approximation: if the substructures inside a structure
are smaller (and therefore more tightly bound), the process of their
disruption by tidal forces and mergers will be much slower and less
efficient.  Indeed, in the limit that $\gamma_{\rm sc} \rightarrow 3$,
any structure which collapses and virializes will ``see" the
substructures which have collapsed before it essentially as point
particles, and thus stable clustering should become exact in this
case. On the other hand, as $\gamma_{\rm sc}$ decreases, we expect
that the interaction between structures can lead more easily to their
disruption, and in particular that mergers of substructures become
much more probable.

There is a simple consequence of the relation (\ref{Eq-scaling-structure})
for numerical simulations. In an N-body simulation
one can in principle follow the non-linear evolution of 
a range of comoving scales initially in the linear 
regime, between a minimal scale (say $L_1^0$) and a maximal scale 
(say $L_2^0$). In units of the initial interparticle separation, 
$L_1^0$ is fixed and independent of $N$, as it can be taken to 
correspond to a region initially enclosing some fixed minimal number of 
particles. The scale $L_2^0$, on the other hand, 
is bounded above by the box 
size (or some fraction of it).  Thus the ratio of these scales
is, to a reasonable approximation, the same for different
cosmological models if one compares simulations  of
a given particle number $N$. It then follows from 
(\ref{Eq-scaling-structure}) that, if stable clustering is a good
approximation in the non-linear regime, the range of
non-linear scales ($L_1$ to $L_2$) in which it will
be observed will increase monotonically with $n$.  
Thus not only do we expect stable clustering to be potentially 
a better physical approximation as $n$ increases, but we also 
expect that the range of scales over which it can be observed
in simulations  of comparable size to increase as $n$ increases.  
This is a behaviour we have already observed  clearly in previous 
studies of this model in both  one \citep{benhaiem2013exponents} and three 
dimensions \citep{benhaiem2013self}.  We will again draw particular 
attention to it in what follows.

%%%%%%%%%%%%%%%%%%%%%%%%%%%%%%%%%%%%%%%%%%%%%%%%%%%%%%%%%%%%%%%%%%%%
%%%%%%%%%%%%%%%%%%%%%%%%%%%%%%%%%%%%%%%%%%%%%%%%%%%%%%%%%%%%%%%%%%%%

\section{Numerical simulations: methods and results}
\label{Numerical_simulations}

\subsection{Equations of motion} % (fold)
\label{ssub:equations_of_motion}

Dissipationless cosmological N-body simulations (see
e.g., \cite{bertschinger_98, springel2005simulations, dehnen+reed_2011})
solve numerically the equations 
\begin{equation}
	\frac{d^2 {\bf x}_i}{dt^2} +
	2H \frac{d{\bf x}_i}{dt}  = \frac{1}{a^3} \bF_i 
	\label{3d-equations-1}
\end{equation}
where the gravitational force is
\begin{equation}
	\bF_i  =- Gm \sum_{j \neq i}^P
	\frac{{\bf x}_i - {\bf x}_j}{\vert {\bf x}_i - {\bf x}_j \vert^3} 
	W_\varepsilon (\vert {\bf x}_i - {\bf x}_j \vert) \,.
	\label{3d-equations-2}
\end{equation}
In Eqs.(\ref{3d-equations-1}-\ref{3d-equations-2}) ${\bf x}_i$
are the comoving positions of the $i=1...N$ particles of equal mass
$m$, in a cubic box of side $L$, and subject to periodic boundary
conditions (the script `P' indicates that the sum
extends over the infinite copies); $a(t)$ is the appropriate scale
factor for the cosmology considered, and $H(t)={\dot a}/{a}$ is the
Hubble constant. The function $W_\varepsilon$ is a regularization of
the divergence of the force at zero separation --- below a
characteristic scale, $\varepsilon$, which is typically (but not
necessarily) fixed in comoving units.

Changing the time variable to 
\begin{equation}
	\tau=\int \frac{dt}{a^{3/2}}\,, 
	\label{tau-definition}
\end{equation}
Eq.~(\ref{3d-equations-1}) can also be cast as 
 \begin{equation}
	\frac{d^2 {\bf x}_i}{d\tau^2} +
	\Gamma \frac{d{\bf x}_i}{d\tau}  =  \bF_i \,,
	\label{3d-equations-3}
\end{equation}
where, for the EdS model given by Eq.~(\ref{eds-scalefactor}), 
we find
\begin{equation}
	\Gamma = \frac{1}{3 t_0}\,.
		\label{Gamma-a-relation}
\end{equation}
In this time variable the equations of motion are thus equivalent to
those of particles in an infinite non-expanding universe subjected to
a simple fluid damping.

In this representation it is clear that the scale-free property is
common to a one parameter family of such models in which $\Gamma$ is
constant, of which the usual EdS model is just one particular case. It
is this more general class of models that, under the same hypotheses
{leading} to self-similar evolution, also permits a simple
generalization of the analytic prediction for the case of stable
clustering, which we have studied in \cite{benhaiem2013self}.  In this
paper we focus on results only for the usual EdS model.

%%%%%%%%%%%%%%%%%%%%%%%%%%%%%%%%%%%%%%%%%%%%%%%%%%%%%%%%%%%%%%%%%%%%%%%%%

%\subsection{Simulation} % (fold)

\subsection{Simulation Code}
\label{sub:simulation_code}

We have made use of the {\tt Gadget-2} code \citep{gadget} with the
modification described in detail in \cite{benhaiem2013exponents} to
simulate the equations as cast in Eqs.~(\ref{3d-equations-3}). We thus
have used the static version of the code (i.e. for a non-expanding
system in periodic boundary conditions), and modified the
time-integration scheme, keeping the original ``Kick-Drift-Kick''
structure of the leap-frog algorithm and just modifying appropriately
the ``Kick'' and ``Drift'' operation to include the damping term. The
structure of the code is otherwise unchanged. 

Tests of this code, notably comparison between results obtained with
it for the EdS case and those obtained using the standard {\tt Gadget-2} 
integration of this cosmology, for the same initial condition and numerical 
parameters, have been discussed at length in \cite{benhaiem2013exponents}.  
Some minor visual differences between such pairs of simulations are observed. 
This is to be expected as these are two different numerical integrations of 
the evolution of a chaotic  dynamics. The statistical properties of the final 
configurations, which is what we measure and study in the present paper, 
are, however, in excellent agreement in the two simulations.

%%%%%%%%%%%%%%%%%%%%%%%%%%%%%%%%%%%%%%%%%%%%%%%%%%%%%%%%%%%%%%%%%%%%%%%%%

\subsection{Simulation parameters}
\label{Simulation parameters}
	
The N body method  introduces in general, {\it three} 
unphysical length scales: the force softening scale $\varepsilon$, 
the mean interparticle separation $\Lambda=n_0^{-1/3}$ 
(where $n_0$ is the mean particle density) and the side
of the periodic box $L$. In a realistic cosmological model (e.g. 
$\Lambda CDM$) there are characteristic physical scales, 
and these three unphysical scales may then be specified in 
physical units, and can be varied independently of one another. 
Usually $\varepsilon$ is referred to as the {\it spatial resolution} 
and given typically in $kpc$,  $\Lambda$ is specified by the 
particle mass ($=\rho_0 \Lambda^3$, where $\rho_0$ is the
mean comoving mass density) and given typically in solar masses,  
and thus referred to as the {\it mass resolution}, and finally $L$, the box size, is 
given typically in Mpc. For a scale-free simulation there is no 
length scale other than the non-linearity scale (defined e.g. 
as the scale at which the mass fluctuations in a sphere are unity).
The standard method of setting up initial conditions (see below) 
fixes this scale relative to the initial interparticle separation
at the initial time, and there are thus in practice then just 
two dimensionless parameters: the ratio $\frac{\varepsilon}{\Lambda}$, 
which we will refer to here simply as the {\it resolution} of the simulation, 
and $N$, the particle number (and $L=N^{1/3} {\Lambda}$). 

We use the version of {\tt Gadget-2} with a force smoothing 
parameter $\varepsilon$ which is fixed in comoving coordinates 
(as is common practice in large volume cosmological simulations, 
and, in particular, in almost all studies of scale-free models). The 
force smoothing in {\tt Gadget-2} is implemented with a spline 
interpolation between the exact gravitational force, at $2.8 \varepsilon$, 
to a vanishing force at zero separation. As anticipated above, we perform 
all of our simulations for two values of the smoothing length: 
$\varepsilon=0.064 \Lambda$, as in \cite{smith2003stable}, and 
$\varepsilon=0.01 \Lambda$, as we have considered in \cite{benhaiem2013self}.
These are 
our ``low resolution" and ``high resolution" simulations respectively.
We consider power law initial conditions with exponents $n = -2,-1,0$.
Table~\ref{Table1} gives the associated predicted stable clustering
exponent $\gsc$, as well as other parameters characterising the
initial amplitude and duration of the simulations which we will 
explain below. 

All of the results reported in this paper are for simulations 
with $N=256^3$. Results for identical simulation parameters
(and values of $\varepsilon/\Lambda$) but with $N=128^3$ 
and $N=64^3$, have been reported in \cite{benhaiem2013self}.
Such simulations are identical up to effects associated with
the differing finite box sizes, which are expected only to
become important from the time that non-linearities
develop at scales approaching the box size. We have
verified that this is indeed the case, and thus that all
results reported here for $N=256^3$ are, other than
finite box size effects (which we detect as deviations
at large scale from self-similarity) independent of
$N$.

One of the central points in our analysis will be the 
differences between these low and high resolution 
simulations. Indeed the question of the optimal 
choice  of $\varepsilon/\Lambda$ is a very important
one for cosmological simulation using the N body 
method which has been widely discussed in the 
literature. While decreasing it can potentially increase
the range of scales in which clustering is resolved,
doing so may compromise the accuracy with which
the $N$ body system can represent the desired 
collisionless limit 
(for a discussion see e.g. \cite{splinter_1998, knebe_etal_2000, romeo_etal_2008, discreteness3_mjbm}).
We will return to discuss this issue at some length 
in the final section. For the present we underline 
that any non-collisional effects arising from the use
of a small smoothing (and, in particular,  those associated 
with two body collisions) should be detected as deviations 
from self-similarity, if they are present at a level which 
significantly affects the quantities we measure. 

From the numerical point of view the use a smaller $\varepsilon$ 
than usually employed in cosmological simulations (exemplified by
the smoothing used e.g. in \cite{smith2003stable}) implies
greater numerical cost, as smaller smoothing requires smaller time
steps to integrate accurately --- and maintain accurate energy
conservation --- on those trajectories on which the forces become
dominated by a single nearby particle (for a discussion, see
e.g. \cite{knebe_etal_2000, joyce+syloslabini_2012,syloslabini_2013}).  
We have performed simple convergence tests as well as tests of 
energy conservation using the so-called Layzer-Irvine equation (see
\cite{benhaiem2013self,joyce+syloslabini_2012,syloslabini_2013} for details). 
We have adopted finally the fiducial recommended
value for the parameter controlling force accuracy, but significantly
more stringent values for the time-stepping parameters as we have
found these to give considerable improvement in the tests using
the Layzer-Irvine equation~\footnote{Specifically we have taken
  {\tt ErrTolIntAccuracy}=0.001, {\tt MaxRMSDisplacementFac}=0.1 and
  {\tt MaxSizeTimestep}=0.01, values which are smaller (by factors of
  $25$ for the first, and $2.5$ for the two others) than the values
  suggested in the  {\tt Gadget-2}  userguide and treated as ``fiducial" in the
  literature (and e.g., in \cite{smith2003stable}). For the force accuracy
  we have taken {\tt ErrTolForceAcc}=0.005 which is a typical fiducial
  value.}. Finally, as underlined, we note again that self-similarity
 in principle robustly tests also for such non-physical effects.

	\begin{table}
	\begin{tabular}{c||c|c|c|c|c|c|c|}
	 $n$ & $\gsc$ &  $\Delta^2_L(k_{N},a=1)$ & $a_f$ & $\Delta^2_L(k_{b},a_f)$  \\
	\hline
	\hline
	0 & 1.80 & 0.94   & 54.6 & 0.0857 \\
	-1 &  1.50 & 0.06  & 54.6 & 0.179 \\
	-2 &  1.00 & 0.03  & 20.1 &  0.196 \\
	\hline
	\hline
	\end{tabular}
	\caption{Parameters characterizing our simulations with $N=256^3$ particles.
	$n$ is the exponent of the initial PS, and $\gamma_{sc}$ the
	exponent characterizing the decay of the CF in the stable clustering
	hypothesis. The quantities  $\Delta^2_L(k_N,t_s=0)$ and $\Delta^2_L(k_{b},t_s^f)$
	characterize the initial and final amplitudes of the fluctuations.
	Note that we use units in which the box size is unity 
	so that wavenumber $k_b$ of the fundamental mode  
	is equal to $2\pi$.}
	\label{Table1}
	\end{table}

To compute the power spectrum, we use the very accurate estimator of
\cite{colombi2009accurate}, and for the correlation function $\xi(x)$,
we used the so-called ``full-shell" estimator which directly counts the
particles in spherical shells centred on a chosen number $N_S$
of randomly selected particles. Our results below are obtained with
$N_S=10^5$. Tests with a larger number and/or different distribution 
of the centres sampled have been performed and we found our results 
to be very stable.Unlike \cite{smith2003stable} we do not attempt 
to apply any corrections to take into account discreteness effects:
again, any such effects should in principle be detectable using the
tests for self-similarity which we focus on here.

%%%%%%%%%%%%%%%%%%%%%%%%%%%%%%%%%%%%%%%%%%%%%%%%%%%%%%%%%%%%%%%%%%%%%%%%%

\subsection{Initial conditions and duration of simulations}

We generate our initial conditions using the code {\it mpgrafic}
\citep{prunet2008initial} which employs the standard method used in
cosmological simulations (see e.g. \cite{bertschingercode,discreteness1_mjbm}): particles, initially on a simple cubic
lattice, are subjected to a displacement field generated as a
sum of independent Gaussian variables in reciprocal space with
variance determined by the desired linear power spectrum, and
including all modes up to the Nyquist frequency $k_N=\pi/\Lambda$
(i.e. we sum over $\vec{k}$ such that each component $k_i \in [-k_N,
  k_N]$.) The initial velocities are then fixed simply by imposing the
growing mode of linear theory using the Zeldovich approximation.

We take an initial power spectrum $P_L(k, a_0)=A_0 k^n$. Following
standard practice, we characterize the initial amplitude by
specifying the value of \[\Delta_L^2 (k_N)= \frac{A_0 k_N^{3+n}}{2
  \pi^2} \;,\] which is (approximately) the normalized mass variance
in a Gaussian sphere of radius $\Lambda$.  In fixing the initial
amplitude of our simulations as given in Table~\ref{Table1}, we have
used as guidance the previous work of \cite{jain+bertschinger_1998}
and of \cite{knollmann_etal_2008} which report tests showing that
self-similarity is recovered better for the cases with smaller $n$ if
low amplitudes are used.

Also given in Table~\ref{Table1} are the final times $a_f$ considered
for our analysis in each of the simulations, and the corresponding
values of the linear theory extrapolated amplitude $\Delta_L^2 (k_b,
a_f)$ at the fundamental mode of the periodic box $k_b=2\pi/L$.  
In the cases $n=-2$ and $n=-1$ our simulations thus extend to times
when the normalized mass variance in a Gaussian sphere of order the
size of the box is no longer much smaller than unity, and one would
expect this to lead to significant finite size effects.  Such effects at
large scales will indeed be detected below using tests for
self-similarity.

\section{Results} 
\label{sec:results}

We divide our analysis into three parts. Firstly,
in Sect.\ref{ssub:effects_of_resolution_on_the_correlation_function},
we compare our results for the CF and PS in the low and high
resolution simulations. 
We then consider, in Sect.\ref{ssub:testing}, what conclusions
concerning the validity of stable clustering may be drawn from 
this comparison.  Finally, in Sect.\ref{sub:self_similarity}, 
we show how, by testing for self-similarity, we can robustly 
determine the range of scales in which the resolved clustering 
is physical. This allows us to draw much stronger conclusions both 
about the limits {on the} resolution {and on} the validity of the predictions derived from
the stable clustering hypothesis.

 %%%%%%%%%%%%%%%%%%%%%%%%%%%%%%%%%%%%%%%%

\subsection{Resolution limits arising from force smoothing} 
\label{ssub:effects_of_resolution_on_the_correlation_function}

\begin{figure*}
		\centering\resizebox{19cm}{!}{\includegraphics[]{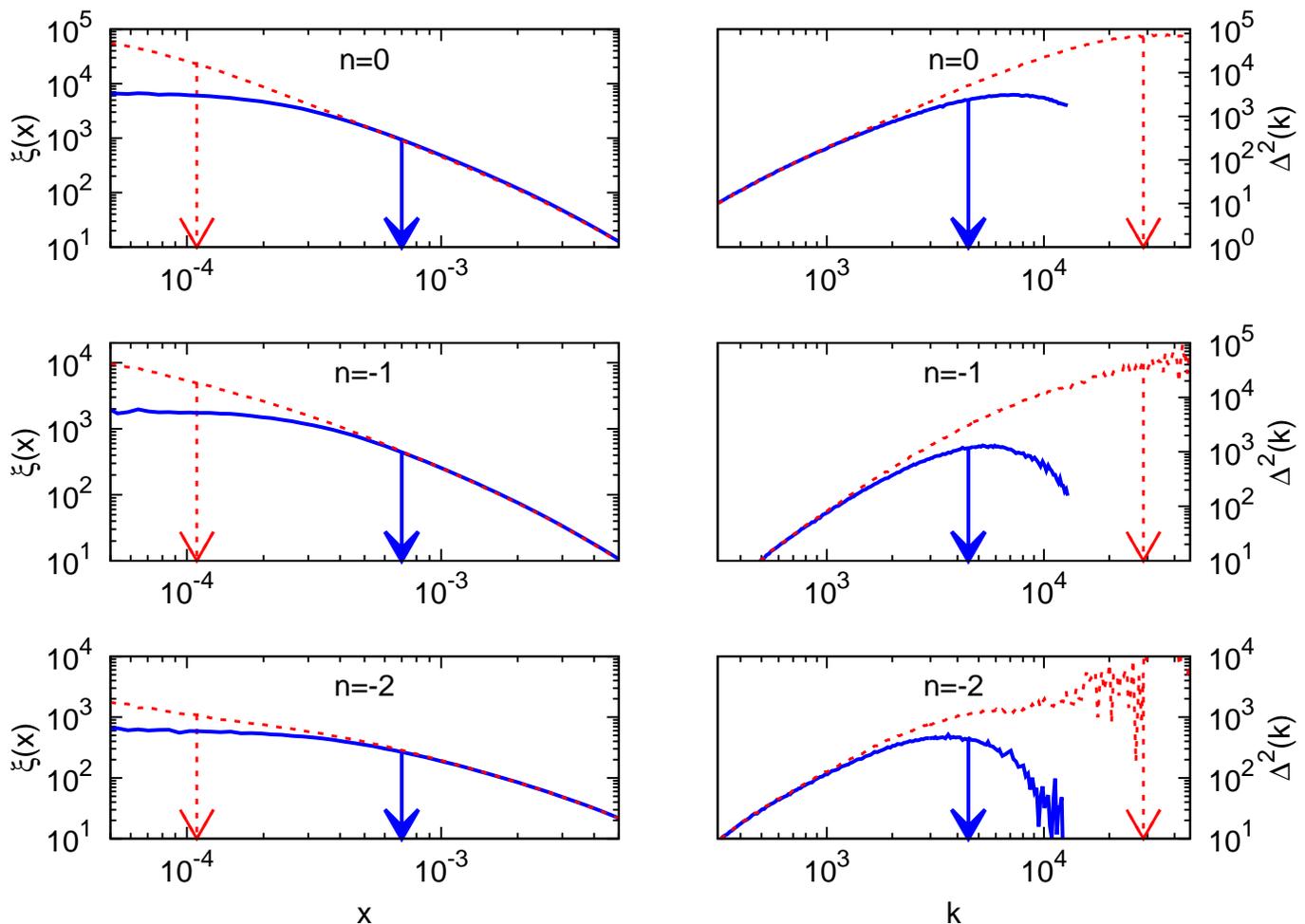} }
		\caption{Two point CF (left panels) and dimensionless
                  PS (right panels) at the final times in the high resolution
                  (dashed red line) and the low resolution (solid blue line) simulations, 
                   for each of the three scale free
                  models: $n=0$ (top), $n=-1$ (middle), $n=-2$
                  (bottom). The vertical lines with arrows indicate
                  the scale $2.8 \varepsilon$ in the left panels, and
                  $\pi/(2.8 \varepsilon)$ in the right panels (solid blue
                  for low resolution, dashed red for high resolution).
                  }
		\label{multiplot-Fig1}
\end{figure*}

We begin by comparing the results of our pairs of low and high
resolution simulations. Shown in Fig.~\ref{multiplot-Fig1} are the CF
(left panels) and the dimensionless power spectrum $\Delta^2(k)$
(right panels) for the three different values of $n$. In particular,
each panel shows the results for both the low (solid blue line)
and high (dashed red line)
resolution simulations.  The blue (red) vertical solid(dashed) line indicates the
scale $2.8 \varepsilon$ in the plots of the CF, and $\pi/(2.8 \varepsilon)$
in the plots of $\Delta^2$, {respectively} for the low (high) resolution
simulation. In these plots we have excluded the very largest scales
(i.e. smallest $k$) at which the differences between the low and high
resolution simulations are, as one would expect, negligibly small.

These plots show very good agreement between the CF in the 
high and low resolution simulations down to the smoothing scale 
of the latter, i.e., down to the scale at which the two body force 
is the same in the two simulations. This behaviour is in agreement 
with what one would naively expect: clustering is apparently  well 
resolved down to the scale at which the force law deviates from the exact
Newtonian force.  Below this scale clustering is clearly
always larger in the higher resolution simulation, so the effect
of the smoothing appears to be simply to suppress clustering 
below the scales at which the force is smaller than its true (Newtonian)
value. We underline, nevertheless, that these results alone
to not allow us to conclude definitively that the clustering in either 
case above the smoothing scale represents accurately the 
physical (continuum) limit. Only the analysis using self-similarity 
described below will allow us to address this question.

The plots of $\Delta^2(k)$ show a distinctly different behaviour: the high and 
low resolution simulations strikingly differ already at a much smaller wavenumber
--- about a factor of four --- than the naively estimated characteristic scale 
for the smoothing. The explanation for this behaviour is again simple: 
the clustering signal at a given wavenumber $k$ picks up contributions
from a range of real space scales around $\pi/k$, and in particular
it can receive significant contributions from scales well below
$\pi/k$.  Indeed, since the CF shows a clustering modified
by smoothing at scales below $2.8 \varepsilon$, its 
Fourier transform is modified over a range of $k$ 
which extends well below $\pi/(2.8 \varepsilon)$. As the force smoothing 
is applied in real space, it is natural for it to have a localised effect on 
clustering in real space, while in reciprocal space its effects on clustering are
``dispersed" into a broader range of wavenumbers, leading in the
present case to very large effects on the PS at {reciprocal space scales 
almost an order of magnitude smaller} than the naively estimated scale.
Again we note that, just as for the CF, this analysis does not allow
us to infer the range of wavenumber in which the measured clustering
in either simulation accurately represents the physical (continuum) limit.

\begin{figure*}
		\centering\resizebox{19cm}{!}{\includegraphics[]{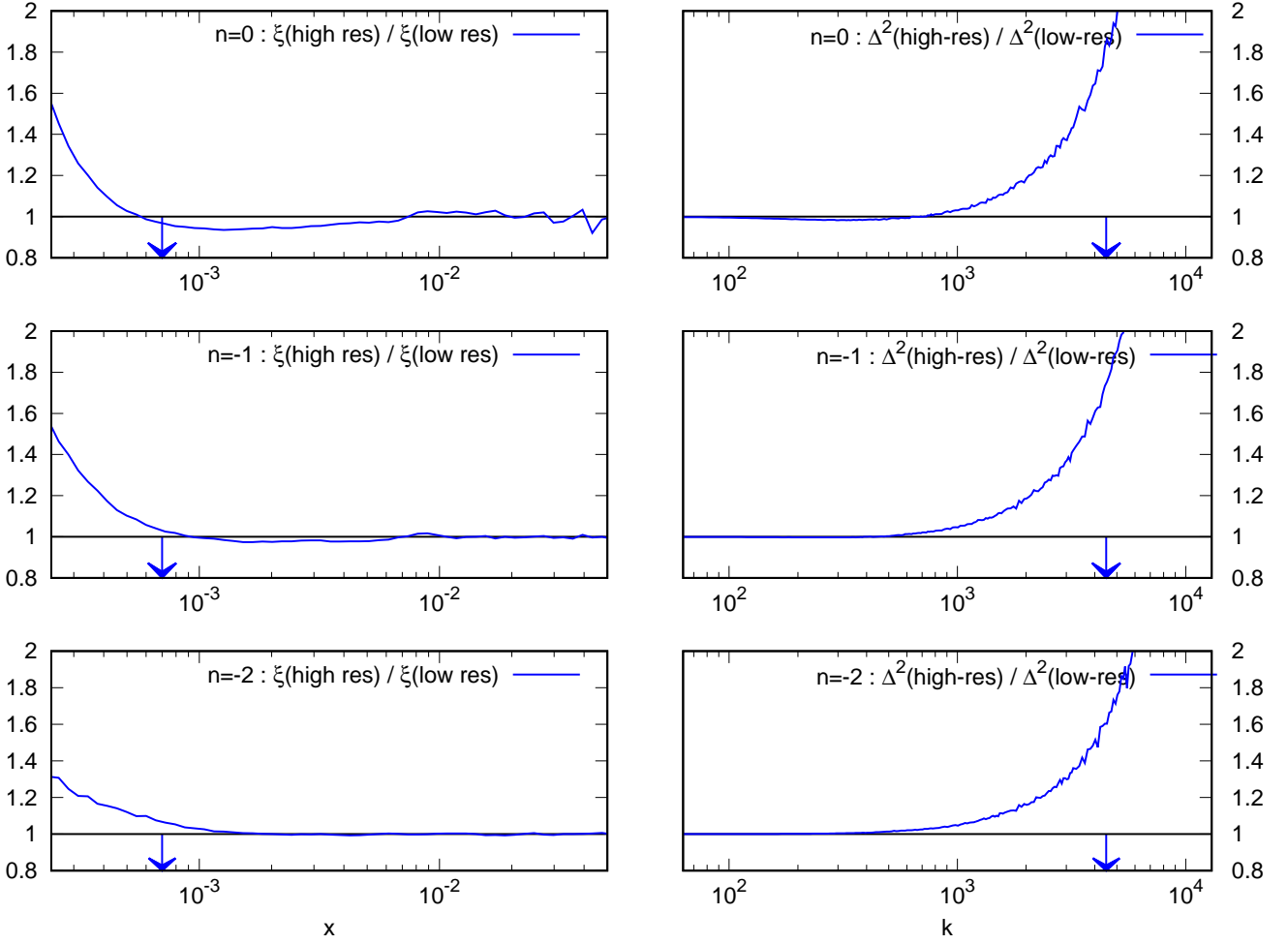} }
		\caption{ Ratio between the high and low resolution simulation 
		of the CF (left) and PS (right). The arrow in each plot gives the force  
		smoothing length of the low resolution simulation.
		}
		\label{fig-RATIO}
\end{figure*}

Fig.~\ref{fig-RATIO} shows the ratio between the measured quantities
in each pair of high and low resolution simulation (CF in left panels,
$\Delta^2(k)$ in right panels): this allows a more detailed
quantitative comparison of these results. The {vertical line with
  arrow} corresponds to the smoothing scale for the low resolution
simulation. The conclusions drawn above concerning the differences
between real and reciprocal space are very clearly visible. However,
we can also note subtle differences as a function of $n$ for both
quantities. In particular, the effect of the resolution on the
CF shows an apparent trend as a function of $n$: for $n=-2$ the 
higher resolution leads to an increased clustering at all scales,
starting from a scale above $2.8 \varepsilon$, while in
the case $n=-1$, and more markedly in the case $n=0$,
such an increase of clustering is seen at smaller scales,
but their is also a small decrease at larger scales. We will
return to this detail further below, as we will see evidence
that it is related to the fact that, in simulations at
fixed $N$,  stable clustering extends to smaller scales 
for larger $n$, as outlined in the discussion at the end of 
Sect. \ref{sub:the_relative_size_of_structures}.

%%%%%%%%%%%%%%%%%%%%%%%%% 

\subsection{Testing the stable clustering prediction, without self-similarity} 
\label{ssub:testing} 

Before turning to our analysis using self-similarity, which will
allow us to determine precisely which part of the clustering signals
at the final time can be assumed to resolve the physical limit,  
let us consider the compatibility of the results above with
the validity of stable clustering in the non-linear regime. As we have 
recalled, the stable clustering hypothesis predicts a power law 
behaviour of these quantities in the strongly non-linear regime, i.e., starting
from scales at which structures may be expected, theoretically, to be
well approximated as virialized. As we will see below (and in
agreement with previous studies) such behaviour is observed starting 
from $\xi \sim 100$ and $\Delta^2 (k) \sim 100$, consistently with 
what would be estimated from the naive spherical collapse 
model (see, e.g., \cite{peebles}). Further, if stable clustering 
is a good approximation, we would expect this always to be the
case in a limited range of scale, corresponding to some 
finite duration after virialization. Thus we would expect that
stable clustering will break down always at asymptotically 
small scales (compared to the non-linearity scale). In seeking to 
test the validity of the predictions of stable clustering one can 
thus envisage three possible behaviours for the CF (or PS):

\begin{itemize}
\item (B1) The CF (or PS) is in good agreement with the predicted 
behaviour  over the full range of resolved highly non-linear scales;
\item (B2) The CF (or PS) is consistent with the predicted behaviour
  over part of the range of resolved highly non-linear scales, but a
  break from the predicted behaviour is detected at a
  sufficiently small scale;
\item (B3) The CF (or PS) is consistent nowhere with the 
predicted behaviour in  the range of resolved highly non-linear scales
\end{itemize}

Note that one could consistently reach different conclusions for
the CF and PS. For instance (B2) could hold for the CF and (B3) for the PS:
if  for the former the range of scale in which the predicted power law behaviour 
is found is very limited, the PS may never be well approximated by 
the ``pure" stable clustering behaviour.

\begin{figure*}
\centering\resizebox{17cm}{!}{\includegraphics[]{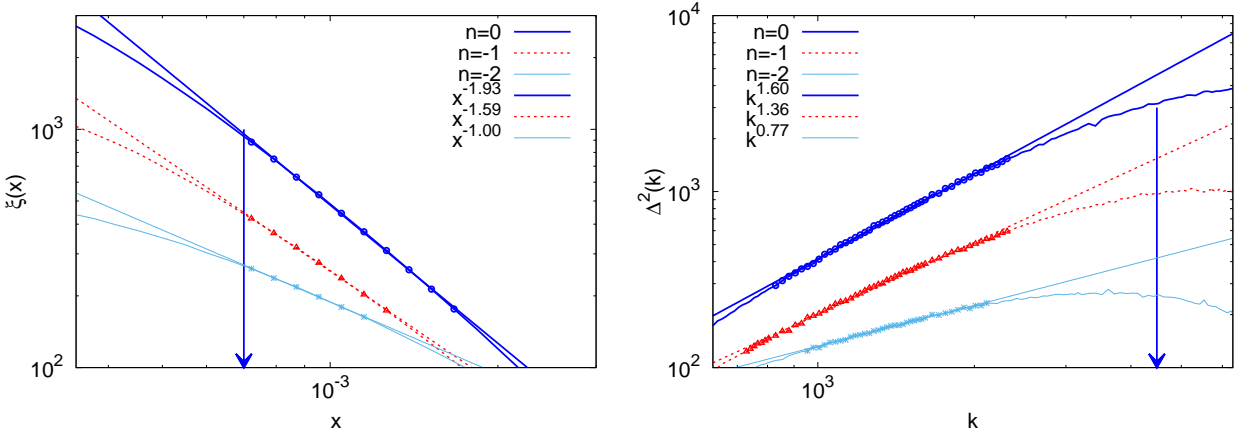}
}
		\caption{CF (left panel) and dimensionless PS (right panel) measured at the
		final time for each of the three low resolution simulations. The blue (thick) solid line is the
                  result for $n=0$, the red (dashed) line is for
                  $n=-1$, and the cyan (thin) solid line for $n=-2$.
                  In each case also shown is a best-fit power law to the discrete data points 
                  plotted, in the range of each curve admitting such a fit by eye.}
		\label{fig-powerlaw-lowrez}
\end{figure*}

\begin{figure*}
  \centering\resizebox{17cm}{!}{\includegraphics[]{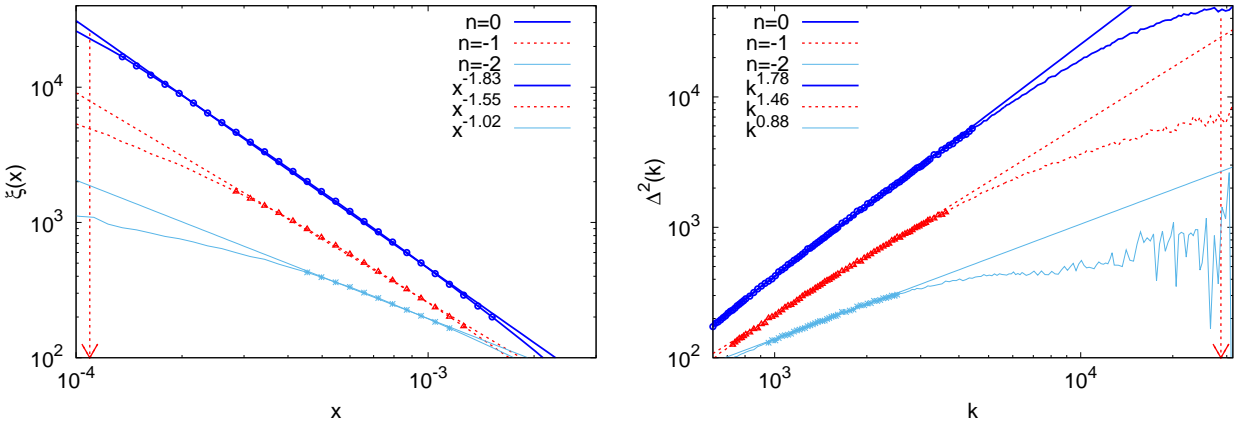}
  }
  \caption{ 
  Same as Fig.~\ref{fig-powerlaw-lowrez} but for the high
    resolution simulations.  
    }
		\label{fig-powerlaw-highrez}
	\end{figure*}

Examining again the plots in Fig.~\ref{multiplot-Fig1} we observe that 
each of the measured CF and PS 
(both { in the high and low resolution simulation}) show, in 
the highly non-linear range, a behaviour which can, by eye, be fitted 
with a power law in a limited range.
This range depends strongly both on the model (i.e. on $n$) and on 
the force smoothing. We find that it extends over a decade in the high
resolution simulation for $n=0$, but barely more than a factor of 
two for $n=-2$. 
Shown in Figs.~\ref{fig-powerlaw-lowrez}-\ref{fig-powerlaw-highrez} are
the best fits to a simple power law obtained in each CF and PS in the
regions which, by eye, appear to potentially admit such a fit.  We
recall that the predicted behaviours for stable clustering are $\xi
(r) \sim r^{-\gamma_{sc}}$ and $\Delta^2 (k) \sim k^{\gamma_{sc}}$
with $\gamma_{sc}=1.8$ for $n=0$, $\gamma_{sc}=1.5$ for $n=-1$ and
$\gamma_{sc}=1$ for $n=-2$.  Thus we observe that {\it the best fit
  exponents for the CF in the high resolution simulations are in
  excellent agreement (within a few percent) with those predicted by the
  stable clustering hypothesis}. {Further,  performing the
  same procedure  on the CF measured in the low resolution simulations,
  we obtain exponents which are again very consistent with those
  predicted by stable clustering, albeit within somewhat larger error 
  bars associated with the more limited range of scale of the fits}.

On the other hand, the exponents obtained in the same way for the PS
are (i) in each simulation systematically smaller than those obtained
with the fit to the CF, and (ii) significantly different in the low
resolution and high resolution simulations.  In particular, the
exponents obtained from the low resolution simulation are considerably
smaller than those obtained by fitting the CF for each case. We
note that these latter values for the effective exponents of the PS
are close to those found by \cite{smith2003stable} ($\gamma=1.49,
1.26, 0.77$ for $n=-2,-1,0$ respectively).  In principle, the results
discussed in the previous section suggest that  the latter
are incorrect, as the lower fitted exponent is clearly a result of a
suppression of power compared to that in the higher resolution
simulation. Given, however, that we have no certainty that the
higher resolution simulation is itself converged in this range,
we cannot draw a definitive conclusion without using
the criterion of self-similarity, as we will describe below.

In summary for the CF a good fit to the stable clustering exponent 
appears to be valid in some range in all simulations, and thus
(B3) appears to be excluded.  In the low resolution simulations the lower 
cut-off to this fit lies in all cases around the smoothing scale ($2.8 \varepsilon$), 
and therefore at the strict lower bound of our spatial resolution. 
Thus, from the low resolution simulations, the conclusion we draw for
all three models is (B1), assuming only that the clustering
represents the physical limit at least in some part of this spatial range.
For the high resolution simulations the same is not true: while
for the case $n=0$ the power law now extends again down to very close
to the smoothing, this is definitely not the case for either $n=-1$ or
$n=-2$.  For $n=-1$ the range of the power law fit extends to a
slightly smaller scale, and the CF bends away to a
much shallower behaviour well above the smoothing length; for $n=-2$
the lower cut-off to the power law fit barely changes and the
CF over the scale resolved with the smaller
smoothing is everywhere well below the extrapolated power law.

Thus, for the cases $n=-1$ and $n=-2$ our high resolution simulations
include a region where the CF is clearly not well
described by the stable clustering prediction. If the clustering in
this region is indeed resolved, i.e.  representative of the continuum
physical model, then our conclusion is (B2). If on the other hand this
is not the case then the correct conclusion is (B1).

As we now discuss the criterion of self-similarity gives us a tool to
answer the crucial question as to what the range is in which the
clustering can be assumed to be physical, for both the CF and PS.

	
%%%%%%%%%%%%%%%%%%%%%%%%%%%%%%%%%%%%%%%%%%%%%%%%%%%%%
%%%%%%%%%%%%%%%%%%%%%%%%%%%%%%%%%%%%%%%%%%%%%%%%%%%%%
%%%%%%%%%%%%%%%%%%%%%%%%%%%%%%%%%%%%%%%%%%%%%%%%%%%%%	

\subsection{Self-similarity tests} % (fold)
\label{sub:self_similarity}

To test for the self-similarity at any given time of a given quantity
--- here the CF or the PS --- we need to simply compare it to its
value at a later or an earlier time in the appropriately rescaled
coordinates, or in more general at a sequence of times. For our
purposes here it will be sufficient to compare the CF and the PS at
the final output time ($a_f$) with the rescaled quantity at a short
time before (but sufficiently long so that there is significant
evolution of the clustering).  We thus compare the CF (and the PS) at
the final scale factor $a_f$ with a slightly smaller scale factor
$a_{f, -}$, chosen so that the rescaling of length scales in all cases
is a factor of $1.5$, i.e., $R_s(a_f)=(1.5) R_s(a_{f,-} )$.  This
corresponds to {${a_{f}}/{a_{f,-}} \approx 2.25, 1.5, 1.23$} for
$n=0,-1,-2$ respectively.

\begin{figure*}
		\centering\resizebox{19cm}{!}{\includegraphics[]{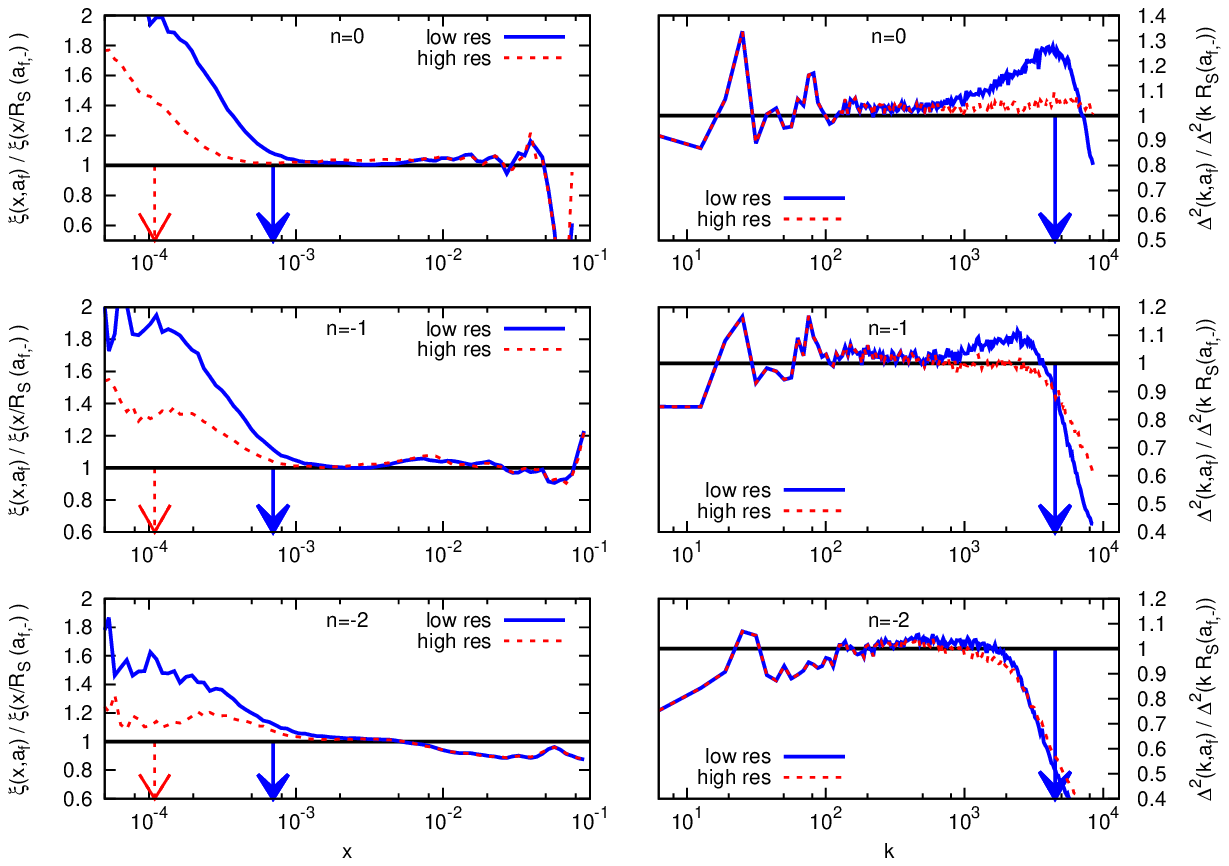}
                }
		\caption{Self-similarity tests: ratio of CF (left
                  panels) and dimensionless PS (right panels) at two
                  scales factors, $a_f$ and $a_{f, -}$ (see text for definition), 
                  where each function is rescaled at the latter scale factor 
                  according to self-similar scaling. From top ($n=0$) to 
                  bottom ($n=-2$), each panel shows result for both high 
                  and low resolution simulation. The arrows in the CF plot
                  indicate the length $2.8 \varepsilon$ for the high and low 
                  resolution simulation, while in the PS plot only the length
                   $\pi/(2.8 \varepsilon)$ for the low resolution falls within the 
                   plotted range.}
		\label{fig-SS}
	\end{figure*}

Shown in Fig.~\ref{fig-SS} are the resulting ratios for the CF (left
panels) and $\Delta^2(k)$ (right panels) for both high and low
resolution simulations for each of the three models. The range of
scales in which this ratio is close to unity corresponds to the range
in which the clustering signal, at both $a_{f}$ and $a_{f,-}$, is
clearly self-similar to a good approximation.  Given that a fuller
analysis shows that the lower cut-off to self-similarity in comoving
coordinates decreases monotonically in time, we can in fact take the
lower cut-off to self-similarity inferred to be that at $a_{f,-}$,
while the lower cut-off at $a_{f}$ may be slightly smaller, and, more
specifically, will be smaller by a factor {${a_{f,-}}/{a_{f}}$} if
stable clustering applies at these scales.

The data in Fig.~\ref{fig-SS} show clearly that the clustering in all
simulations breaks self-similarity both at {large and small}
scales, while in an intermediate range self-similarity is obtained to
a good approximation. This is very much as expected, and shows the
power of the method to detect non-physical effects: at large scales
self-similarity is violated due to finite box size effects, and at
small scales due to effects arising from the ultraviolet cutoffs,
i.e. the force smoothing and the finite particle number sampling of
the continuous density field. That the two asymptotic regimes are
reasonably well separated means that there is in principle no mixing
of the two kinds of effects, and this is further confirmed by the fact
that the deviations at large scales are completely insensitive notably
to the force resolution. 

{We note} that the differences in $\Delta^2(k)$ at small $k$ come
predominantly from the sparse mode sampling, while in the CF they
arise from the sparseness noise in the estimator at very large
scales. Further we observe that the deviations from self-similarity at
large scales are much more marked and extend further, {in
  particular}, for the case $n=-2$ .  This is due, as has been
previously documented (see in particular
\cite{jain+bertschinger_1998}), to the stronger coupling to long
wavelength modes characteristics of lower $n$ spectra, which makes the
detection of self-similarity more difficult in this case.

In the behaviour at small scales (large wavenumber for PS) 
in Fig.~\ref{fig-SS} we observe, in contrast, an evident and
very non-trivial dependence on the force smoothing.  Comparing the
results for the high and low resolution simulations, we see clearly,
both in real space and reciprocal space, that, for $n=0$, 
the range in which self-similarity holds to a very good approximation
is extended  {\it considerably} when the resolution is increases 
(i.e. the softening decreased). For $n=-1$ a similar effect is seen, 
albeit less pronounced, while  for $n=-2$ the scale at which self-similarity 
breaks down does not  appear to move significantly at all (in either space).
These behaviours appear to reflect closely those we observed above 
for the bending in the CF and PS, and suggest that the location of 
this bending may be strongly correlated to the breakdown of 
self-similarity.  Making now, for the CF and each $n$, a quantitative comparison 
between the lower cut-off to self-similarity (estimated, e.g., as that at which the ratio
deviates by more than $10\%$ from unity) and the length scale at which
the bending of CF away from a power law behaviour consistent with
stable clustering was observed above, we see that in all cases the
former scale is in fact substantially greater than the latter. Further if we
assume that stable clustering does indeed hold around these scales and
thus infer that the scale at which self-similarity breaks at $a_f$
is indeed smaller by the factor ${a_{f,-}}/{a_{f}}$, we find that,
in all cases, the scale at which self-similarity breaks coincides very
closely with the scale down to which the stable clustering
approximation describes well the CF.

The explanation of this very strong correlation of these scales
{is, we believe, simply} the following: {\it the mechanism which
 propagates self-similarity to smaller comoving scales in the highly
 non-linear regime is the stability of the clustering} (in the
corresponding range of space and time scales).  And, conversely, when
the clustering is not stable --- i.e., no longer well described as the
stable evolution of the virialized collapsed initial over densities
--- there are strong deviations from self-similarity.
Indeed such an association of the two scales is very plausible:
the breakdown of the stable clustering hypothesis is associated 
with a change in the physical processes characterising the 
evolution of clustering. In particular {it is} associated with the 
interaction of structures and even {with   their} merging.  Both 
the occurrence and outcome of such processes can be expected 
to be sensitive to the non-physical ultraviolet scales associated with 
discretisation.  In particular it is clear that the smallest structures 
which form in simulations of this kind are subject to discretisation 
effects which can propagate to progressively larger scales 
as structures merge and interact.
 
This hypothesis is further born out by the dependence on $n$ of the
scale of the break from stable clustering: as explained in
Section~\ref{sub:the_relative_size_of_structures}, we expect, in
simulations that follow, as here, a comparable range of initial
comoving scales from the linear into the highly non-linear regime, the
ratio of the range over which stable clustering can be observed
increases with $n$. This is qualitatively very clearly in
line with what is observed, in the high resolution simulations
(cf. Fig.~\ref{fig-powerlaw-highrez}), while it is obscured in the low
resolution simulations (cf. Fig.~\ref{fig-powerlaw-lowrez}) because of
the larger smoothing scale.

Related to this, we remark also on another interesting feature 
of the results  in Fig.~\ref{fig-SS}, and in particular for the PS
(right panels). Comparing the results for the low resolution
simulation in the three cases, we observe that the wavenumber 
at which a clear deviation from self-similarity develops 
decreases as $n$ increases, i.e., for $n=0$ this
wavenumber is smallest, while for $n=-2$ it is largest.
Furthermore, there is also a marked difference in
the form of the deviation from self-similarity:
for $n=-2$ the deviation at small scales is towards 
a suppression of the power, while for $n=-1$,
and more markedly for $n=0$, there is a 
clear positive ``bump". The fact that these 
bumps vanish in the higher resolution simulations 
shows clearly that they are a result of using 
a smoothing which is larger than the scale
to which the self-similarity can actually 
propagate, through stable clustering, in the 
absence of the smoothing. The reason is simple
to understand: using such a large smoothing,
the clustering mass, rather than becoming 
ever more concentrated at smaller scales, 
gets ``frozen", at a scale of order $\varepsilon$
in direct space. In reciprocal space, which
mixes over a range of small scales, this
excess power is redistributed in a range
of scales above $\pi/\varepsilon$. Thus
we conclude that the use of a smoothing
scale larger than that to which stable
clustering can actually propagate 
clustering in the duration of the simulation
may further degrade the self-similarity
of the PS in particular. 

\subsection{Exponents of self-similar non-linear clustering} % (fold)

We can check now in more detail the degree of agreement with the
stable clustering predictions for the CF and PS, using the constraint
imposed by self-similarity in the high resolution simulations, which
clearly show self-similar behaviour in a broader range for several cases.
The precise range one chooses to fit in is somewhat arbitrary, as it
depends on how large a deviation from self-similarity one chooses to
tolerate. We follow the procedure described in
\cite{benhaiem2013exponents}, performing power law fits to both the CF
and PS between an upper cut-off, chosen by eye where the functions
break from a power law behaviour (corresponding in each case to
$\xi$ or $\Delta^2(k)$ between $100$ and $200$), and a lower cut-off, 
fixed in two different ways: (i) as the scale $x_{SS}$ ($k_{SS}$) at which
the quantities plotted in Fig.~\ref{fig-SS} deviate by less than $10 \%$
from unity, and (ii) as the scale $x_{SS}^{'}$ ($k_{SS}^{'}$) obtained
by extrapolating $x_{ss}$($k_{SS}$) assuming stable clustering to be
valid between $a_{f,-}$ and $a_f$.  We impose also the constraint that
these latter scales are larger (smaller) than {$\varepsilon$} ($5
k_\varepsilon$) for the CF ($PS$) fits. In the case $n=0$ there is only
one fit to the CF because $x_{SS}^\prime < 2.8 \varepsilon$.  The
results of these different fits are reported in Table.\ref{Table2},
along with the values reported by \cite{smith2003stable}.
\begin{table}
\begin{tabular}{c||c|c|c|c|c|c|c|c}
n & $\gamma_{sc}$  &  CF (1)  
&  CF (2) & PS (1) & PS(2) & Smith et al. \\ %\cite{smith2003stable} \\ 
% &  $\overline{\gamma} $ 
\hline
\hline
0   & {\bf 1.80} & 1.87 & 1.81 & 1.76 & &  1.49 \\
\hline
-1  & {\bf 1.50} & 1.62 & 1.56 & 1.44 & 1.39  &1.26 \\
\hline
-2  & {\bf 1.00} & 1.08 & 1.04 & 0.90 & 0.86  &  0.77 \\
\hline
	
\end{tabular}
\caption[ ]{Theoretical stable clustering exponent ($\gamma_{sc}$)
and the corresponding measured exponents for the different $n$
  models, obtained by fitting the strongly non-linear CF and PS in the
  two range of scales described in the text: (1) indicates the more
  restricted range, (2) the extrapolated range. We also show for
  comparison the exponents reported by \cite{smith2003stable}. }
	\label{Table2}
	\end{table}

By modifying marginally the fitted ranges we find best-fits with
exponents varying by of order $5 \%$.  As can be seen 
from Table.\ref{Table2} our measured exponents for the
CF are thus in very good agreement with the stable clustering
predictions, albeit of course with a large error bar for $n=-2$
corresponding to the very limited range of the fitted region. For the
PS the exponents appear to be a little lower than predicted, but the
most likely explanation for this is that the results for the PS are
still not converged. Indeed we observe  in Fig.~\ref{fig-SS} 
that there is a significant deviation from unity in the range
which has been treated as self-similar. 

\subsection{Comparison with other studies} % (fold)
Our results are in clear disagreement with those of
\cite{smith2003stable}, and the explanation is clearly that their fits
have been performed solely with the PS in low resolution runs, and
have been slightly extended into regions in which the power is 
suppressed and self-similarity broken.  For the case $n=0$, we
note that we could have erroneously obtained a lower exponent more
consistent with the value of \cite{smith2003stable} also in the high
resolution simulation by fitting the PS blindly using only the chosen
criterion on the breaking of self-similarity, which in fact indicates
it may extend in this case right up to $k_{\varepsilon}$.

A study of scale-free simulations has also been reported 
by \cite{widrow_etal2009} , for $n \in [-2.5, -1]$, 
and $N$ varying from $32^3$ to as large as $1584^3$. 
A single value of the smoothing length in units
of the interparticle distance is employed in all simulations,
equal to half that  of \cite{smith2003stable} (and our
low resolution simulations) and thus three 
times larger than in our high resolution simulations.
As in  \cite{smith2003stable} only a reciprocal space analysis of the 
PS is considered.  Very significant discrepancies between the 
measured PS and those of  \cite{smith2003stable} are found: 
due to the smaller softening there is measurably more power at 
large $k$, in line with what we have found here.  However, for what 
concerns the slope of the PS at the largest $k$ fitted, the results 
found are very consistent (for the common cases at $n \leq -1$) with 
those of \cite{smith2003stable}, and the conclusions of the
paper regarding the breakdown of stable clustering concord
with those of \cite{smith2003stable}: the ``asymptotic" 
logarithmic slope, labelled $\bar{\mu}$ of the measured 
PS are significantly  different from that predicted by the 
PD fitting formulae (which is constructed to reproduce
the stable clustering prediction at large $k$ ). We believe
the analysis provided in this paper on this crucial point has 
the same  essential shortcomings as that of  \cite{smith2003stable},
and that this conclusion is not convincingly demonstrated.
More specifically, it is based  on the observed tendency 
(cf. the lower panels of Figure 6 
(for $n=-1$)  and Figure 7 (for $n=-2$) in \cite{widrow_etal2009})
of the measured slope $\bar{\mu}$  towards just slightly larger values 
than those predicted by stable clustering over last decade in $k$: 
for $n=-1$, $\bar{\mu}$ varies between $-1.5$  and $-1.8$
(where the former is the stable clustering value), and for
or $n=-2$, $\bar{\mu}$ varies between $-1.9$  and $-2.2$
 (where the stable clustering value is $-2$).
In these figures it is evident that the different curves (at different times), 
which should overlap if there is self-similarity, in fact show a significant 
dispersion, and at the largest $k$ there is only data for the last time
(and therefore no check of self-similarity).  Given that we
have seen that the PS typically begins to be visibly
suppressed by softening at the corresponding wavenumbers 
(well below the naively estimated scale, $\sim \pi/\varepsilon$), and 
that such a suppression would lead also to a decrease in the fitted 
logarithmic slope.  Only by testing these results (i)  for their
independence of the choice of the smoothing length, 
by doing higher resolution simulations as we have done, 
and (ii) performing the analysis of the CF in real
space,  could one  confidently conclude that these
(small) deviations from the stable clustering prediction
are physical. \cite{widrow_etal2009} have not performed
such a test, and the detailed comparison we have
performed here using such a test on the results 
of  \cite{smith2003stable}, which have turned out to be 
the result of under-estimating the effect of force smoothing 
on the PS, suggest that much care is needed on
this point before definitive conclusions can be
drawn. In short when strong conclusions are drawn
from a very limited range of large $k$ closed
to the inferred resolution limit in $k$, one must
be very sure indeed that this resolution limit
has been very robustly determined.

%%%%%%%%%%%%%%%%%%%%%%%%%%%%%%%%%%%%%%%%%%%%%%%%%%%%%%%%%%%%%%%%%%%%
%%%%%%%%%%%%%%%%%%%%%%%%%%%%%%%%%%%%%%%%%%%%%%%%%%%%%%%%%%%%%%%%%%%%
%%%%%%%%%%%%%%%%%%%%%%%%%%%%%%%%%%%%%%%%%%%%%%%%%%%%%%%%%%%%%%%%%%%%

\section{Conclusions} % (fold)
\label{sec:stable_clustering}
 
We have revisited the study of scale-free models, with a focus on
using them as a tool to understand better what the resolution of
cosmological N-body simulations {truely are, i.e.,} how reliably such
simulations can reproduce the clustering in the continuum physical
limit.

%%%%%%%%%%%%%%%%%%%%%%%%%%%%%%%%%%%%%%%%%%%%%%%%%%%%%%%%%%%%%%%%%%%%

\subsection{Resolution in the strongly non-linear regime}
Our main finding is that the measures of two point
statistics in the strongly non-linear regime of our scale-free 
simulations, represent accurately the physical limit 
only in the range of scales in which stable clustering
remains a good approximation. Indeed we have found a very clear and
robust association between the real-space scale at which the two point
{CF} deviates from the behaviour predicted by stable clustering
and the scale at which self-similarity breaks down. We have explained
that such an association is natural because the breakdown of stable
clustering is indeed associated with physical processes which may
intrinsically be much more sensitive to fluctuations at scales
affected by the ultra-violet cut-offs --- notably the grid scale and
force softening --- introduced by the N-body discretisation.

Let us underline, firstly, that our conclusion is not that 
strongly non-linear physical clustering which is not stable {\it cannot} be
resolved accurately in an N-body simulation, but just that in practice 
it is {\it not} accurately resolved in those we have done, nor in those of
\cite{smith2003stable} and of \cite{widrow_etal2009}, which are fairly typical 
of current cosmological simulations.  Conversely the physical processes such 
as merging which violate the stable clustering approximation are, in our
simulations and those of \cite{smith2003stable}, apparently
polluted by discreteness effects, and the corresponding clustering, which is
measured in some cases over a significant range of scale, cannot be
assumed to represent accurately the physical limit. 

Secondly, we cannot and do not conclude that {\it all} N-body
simulations in the literature of realistic cosmological initial
conditions fail to resolve the regime in which stability of clustering
is not a good approximation. We believe, however, that our results 
place in serious question, at least, the accuracy of all such results. 
As a consequence they place in doubt the accuracy in particular of 
popular phenomenological fits to the strongly non-linear regime 
based on halos models.  Indeed \cite{smith2003stable} is one of the 
reference studies in the literature for such fits (in particular the``halofit" model),
and the fact that we have found its results to be not only
quantitatively, but also qualitatively, incorrect for the case of
scale-free initial conditions logically places in doubt the
correctness of its interpretation of its simulation results for the
case of spectra which are not scale free.  

As we have noted in the introduction, higher resolution
simulations by \cite{takahashi_etal_2012} for these cases have in fact
shown that the results of \cite{smith2003stable} at small scales to be
manifestly resolution dependent.  Our analysis of the scale-free
models leads us to the conclusion that the real limits on resolution
imply that, rather than adjustment of the best fit parameters of the
phenomenological halofit model, it is the correctness of fitting to
any such model breaking stable clustering in the strongly non-linear
regime which should be placed in question.
 
%%%%%%%%%%%%%%%%%%%%%%%%%%%%%%%%%%%%%%%%%%%%%%%%%%%%%%%%%%%%%%%%%%%%

\subsection{Real space vs. reciprocal space analysis}

One important aspect of our analysis is that we studied always in
parallel the two point correlation properties in both real and
reciprocal space. It is very clear from our results that, to
understand the issue of spatial resolution, and also indeed that of
stable clustering, is it absolutely essential to consider carefully
the real space quantities: the physical phenomena are expected to be
characterized fundamentally by real space scales and the mixing of
real space scales in reciprocal space makes it much more difficult to
identify the essential dependencies.  Indeed we believe that the
erroneous conclusion of \cite{smith2003stable} are essentially due to
the use of a k space analysis only.

%%%%%%%%%%%%%%%%%%%%%%%%%%%%%%%%%%%%%%%%%%%%%%%%%%%%%%%%%%%%%%%%%%%%

\subsection{Choice of force smoothing}

As we have noted, the question of what is the optimal smoothing
for an N-body simulation of a cosmological model is an important open one,
and we now summarise what conclusion we draw from our study about
it.

There are two different, but related, aspects to this question of 
optimisation. On the one hand, there is consideration of numerical cost: the
smaller the smoothing, the greater the numerical cost to integrate accurately
the $N$ body system.  On the other hand, the use of a large smoothing
bounds below the length scale which can be resolved, while too
small a value can potentially amplify discrete effects  --- most evidently, 
two body collisions --- which do not represent the physical 
collisionless limit. The question of its optimisation can thus 
be phrased as follows: how small a value of the force smoothing 
should be taken to maximize the range of scales over which 
physical clustering can be accurately simulated? 

Our results show clearly that {\it reducing} force smoothing, down
to the values we have considered, somewhat smaller than those
typically used in cosmological simulation, never decreases the range 
in which non-linear clustering is self-similar (i.e. physical) to a good
approximation, but can, depending on the model, increase this
range.  In other words in no case have we found evidence that using 
higher resolution produces any significant degradation of the lower
resolution result, and can, on the other hand, signicantly  
extend  the range of resolved clustering (most strongly for 
$n=0$). In particular we infer from this 
that any associated  additional two body scattering does not sensibly 
affect the quantities we measure. This is reasonable as we have indeed,
as detailed in Sect. \ref{Simulation parameters}, increased 
numerical accuracy specifically to ensure accurate integration
of the consequent less soft two body collisions (and 
the rate of two body collisionality is in fact only weakly
dependent on $\varepsilon$, remaining finite even
at $\varepsilon=0$). We have, on the other hand, found clear  
evidence that using a force smoothing which is larger
than the scale down to which self-similarity can 
potentially propagate in the duration of the 
simulation (i.e. as seen in the higher resolution
simulation) can lead to a significant degrading
of the results, for the PS in particular.

In summary our results indicate that there is no
apparent reason for using a finite smoothing in cosmological 
simulations other than a consideration of numerical cost:
provided the numerical accuracy is sufficient, we
have not found any evidence of adverse effects
of using a small smoothing.  Such effects may of
course exist, and manifest themselves at yet
smaller values of $\varepsilon/\Lambda$, but 
we have not found them. We note that, for
what regards two body effects, this is quite
consistent with the conclusion of other detailed
studies (e.g. \cite{knebe_etal_2000, joyce+syloslabini_2012}).
Taking   numerical cost into account, our conclusion is then 
that the optimal smoothing for scale-free simulations
--- at least for the  determination of the  two point statistics 
we have  studied --- is that which allows the resolution of 
the scale down to which self-similar clustering 
would propagate in the duration of the simulation
if $\varepsilon$ were zero. In a non scale-free
simulation, the equivalent would be expected 
to be the scale down to which the non-linear
clustering is dominated by the density 
fluctuations initially modelled well in 
the initial conditions.

In any model, if strongly 
non-linear clustering is stable to a good approximation,  
this minimal scale fixing the optimal softening can easily 
be estimated: it is  $\sim L_v^0 (a_v/a_f)$
where $L_v^0$ is the average comoving size of the 
first resolved non-linear structures (containing e.g. $10^2$
particles) when it virializes, at a scale factor 
$a_v$, and $a_f$ is the final scale factor.  Let us consider 
just how this depends, in a given model, on the size of the 
simulation  (i.e. $N$).  In units of the interparticle 
separation $\Lambda$  it just decreases as the inverse 
of the final scale factor (assuming fixed amplitude of 
power at the scale $\Lambda$), which  is 
fixed just by $N$. Specifically, for a scale-free
simulation, assuming simulations are stopped
when the non-linear scale is a fixed fraction
of the box size, we have $a_f \propto N^\frac{3+n}{6}$.
For our $N=256^3$ simulations we have seen
(cf. Fig.~\ref{fig-SS}) that the low resolution value
(used also by \cite{smith2003stable}) appears to 
be close to optimal for the case $n=-2$, but larger 
than the optimal value for the other two cases.  In
the latter cases our results do not allow us to
conclude whether our high resolution values
are optimal either: to do so we would need to simulate
with yet smaller $\varepsilon/\Lambda$ to
see whether we can extend the range of 
measured self-similar clustering. Concerning 
the simulations of \cite{widrow_etal2009}, 
which use an $\varepsilon/\Lambda$ half that
of \cite{smith2003stable}, and $N$ up to
a factor of $4^3$ larger, the resolution 
appears also close to optimal for
$n=-2$ but again significantly larger than
optimal for $n=-1$.

\subsection{Future studies}
 Our final conclusion from the  present study is that further larger,
 studies of scale-free models should be undertaken to try to establish 
 whether the breaking of stable clustering can be unambiguously
 detected in an N-body simulation, ideally  of comparable sizes to the 
 largest simulations currently performed in the community. 
 As we have discussed, such simulations at larger particle number
should be performed over a range of resolution (i.e. values of the
parameter $\varepsilon/\Lambda$) which extends to the
limit in which the range of self-similar clustering observed
becomes independent of its value, i.e., in the case of stable
clustering a resolution high enough to follow the stable
evolution of the first virialized structures through to the
end of the simulation. Unless it can be shown unambiguously in scale-free  
models, using a combined analysis both in real and reciprocal
space, that self-similarity extends into the non-linear region 
where the  predictions of the stable clustering hypothesis
are clearly wrong, we conclude that one can have
 little confidence that realistic cosmological simulations, where the
 test of self-similarity is not available, can in fact accurately
 trace the physical clustering into the same regime.
 We note that large ($N=1024^3$) scale-free 
 simulations with quite high resolution have in fact been performed recently 
 by \cite{diemer+kravtsov_2015}, but analysed only to determine 
 the properties of halos extracted from them and without detailed
 consideration of tests for self-similarity (or indeed tests for 
 stable clustering). In a forthcoming study, 
 using simulations similar to those presented here,   we will 
 also explore in detail the clustering in  scale-free simulations in 
 terms of halo properties, and address in details the question of which 
 of the measured properties in simulations can be shown to be self-similar
 and therefore physical). In particular we will aim to determine which 
 scales are resolved  within the halos, and how their properties are 
 related to that of the CFs.

\bigskip 

{Our numerical simulations have been run on the HPC resources of The
  Institute for Scientific Computing and Simulation financed by Region
  Ile de France and the project Equip@Meso (reference ANR-10-EQPX-
  29-01) overseen by the French National Research Agency (ANR) as part
  of the Investissements d'Avenir program.}

{We are indebted to Bruno Marcos for his collaboration in the first phase
of this project, and specifically for his contribution to the development of
modified version of the Gadget code.  We thank E. Bertschinger, S. Colombi,
B. Diemer,  C. Orban, R. Sheth and J. V$\ddot{\rm a}$liviita for useful conversations 
or remarks.}

%\bibliographystyle{mn2e}
%\bibliography{bibliography}

\setlength{\bibhang}{2.0em}
\setlength\labelwidth{0.0em}

\end{document}